\documentclass[12pt, preprint]{aastex}

\usepackage{emulateapj5}  

\def\einstein{{\it Einstein }}	
\def\ergcm2s{~erg cm$^{-2}$ s$^{-1}$ } 
\def\ergs{~erg s$^{-1}$}		
\def\cmsq{~cm$^{-2}$ }		%

\def\etal{et al.~}		
\def\kms{~km s$^{-1}$}		

\def\msun{~M$_{\odot}$}

\def\deg{$^{\circ}$}
\def\chandra{{\it Chandra }}

\def\x2{$\chi^{2}$}	



\begin{document}

\title{NGC~4261 and NGC~4697: rejuvenated elliptical galaxies}

\author{A. Zezas, L. Hernquist, G. Fabbiano, J. Miller}
\affil{Harvard-Smithsonian Center for Astrophysics, 60 Garden
Street, Cambridge, MA 02138}
\email{azezas@cfa.harvard.edu;lhernquist@cfa.harvard.edu;
pepi@cfa.harvard.edu;jmmiller@cfa.harvard.edu}
\shorttitle{Rejuvenated elliptical galaxies}
\shortauthors{Zezas et al}
\bigskip

\begin{abstract}

\chandra images present evidence for a non-uniform spatial distribution of discrete
X-ray sources in the elliptical galaxy NGC~4261.  This non-uniform
distribution is inconsistent with the optical morphology of NGC~4261
at greater than a 99.9\% confidence level.  Similar evidence is seen in
one more elliptical galaxy (NGC~4697; 98\% confidence level) 
out of five cases we investigated.  NGC~4261 and NGC~4697 have 
old stellar populations  (9-15~Gyrs) and 
fine structure parameters of 1 and 0 respectively,
 suggesting no recent merging activity.  On
the basis of simulations of galaxy interactions, we propose that the
X-ray sources responsible for the non-uniform distribution are
associated with young stellar populations, related to the rejuvenating
fall-back of material in tidal tails onto a relaxed merger remnant, or shock
induced star-formation along the tidal tails.

\end{abstract}


\section{Introduction}

Numerical simulations of galaxy interactions suggest that mergers of
spiral galaxies can lead to the formation of elliptical galaxies
(e.g. Toomre \& Toomre 1972; Barnes 1988, 1992; Hernquist 1992, 1993).
These expectations are supported by optical imaging observations which
show that several elliptical galaxies exhibit shells, ripples, arcs,
counter-rotating cores, or faint tidal tails; structures which are
interpreted as evidence for galaxy interactions or mergers
(e.g. Schweizer \etal 1990).

The availability of high spatial resolution X-ray data allows us to 
investigate these indications using X-ray binary populations
as probes of the star-formation histories of galaxies: 
 High-Mass X-ray binaries (HMXBs) form much more efficiently 
 than  Low Mass X-ray binaries (LMXBs) (Kalogera \& Webbink, 1998;
Portegies Zwart \& Verbunt, 1996), leading to a higher number of X-ray sources per  star in young stellar populations. 
Thus,  using number counts of discrete X-ray sources, we can identify regions of recent or enhanced star-formation within a galaxy. 
 For example, the spatial distribution of LMXBs (which trace the 
old stellar populations) in early-type galaxies is expected to  be
smooth, generally following
the distribution of optical star-light. On the other hand, if there
are any sites of
recent star-formation (e.g. triggered by galaxy interactions)
  they are expected to host  HMXBs, the larger numbers of which
(compared to the LMXBs) may 
result in an overall non-uniform X-ray source spatial distribution.

In this paper we present evidence for a non-uniform spatial
distribution of X-ray sources in NGC~4697, an elliptical galaxy
without any indication for merging activity from optical data, and we
compare these results with similar observations of other early-type
galaxies, finding a second example in NGC~4261. 
 A more detailed investigation of the X-ray source
populations, their degree of non-uniformity and links to other
merger activity diagnostics will be presented in a forthcoming paper.

\section{NGC~4261}

 NGC~4261 is a nearby (35.1~Mpc for
$\rm{H_{o}}$=75$~\rm{km/s}$) E2 type galaxy, belonging to a poor group 
of galaxies (Garcia, 1993) but without  any evidence for
interaction with other members of the group. Optical observations show 
a 20\arcsec ($\sim3.4$~pc) long dust lane along the North-South axis
of the galaxy (Martel \etal 2000 and references therein) and a
spectacular nuclear dust disk ($\sim2\arcsec$; Jaffe \etal 1996, Ferrarese \etal
1996). Its optical isophotes in larger scales have a very strong boxy
morphology (e.g. Peletier \etal 1990). Optical spectroscopy
shows that its dominant stellar population has an age of $\sim15$~Gyr
(Trager \etal 2000).  
This galaxy
also features two prominent radio jets (e.g. Birkinshaw \& Davies
1985, Jones \etal 2000) emanating from an active nucleus. The mass of
the supermassive nuclear black-hole is estimated to be
$(4.9\pm1.0)\times10^{8}M_{\odot}$ (Ferrarese \etal 1996).

\subsection{Observations and Data Analysis}

NGC4261 was observed for 35~ks (OBSID 834; PI M. Birkinshaw) with the
ACIS-S3 CCD on-board the \chandra X-ray
Observatory. The spatial resolution of \chandra
is 0.5$''$ (van Speybroeck \etal 1997) corresponding to a physical
resolution of $\sim80$~pc at the distance on the galaxy.  Although the observation was performed
in 1/2 subarray mode in order to mitigate pile-up on the nuclear
source, the active field of view ($4.1'\times8.3'$) was large enough
to cover the whole galaxy ($\rm{D_{25}=4.07'}$; de Vaucouleurs et al.,
1991).  The initial processing of the data and the results from the
analysis of the nuclear spectrum are presented in Zezas \etal
(2003). In this paper we concentrate on the spatial distribution of
the discrete sources.

We used the {\textit{wavdetect}} algorithm, within the CIAO v2.2 data analysis
suite in order to  search for discrete sources  in four different energy bands: full band (0.3-7.0~keV),
soft band (0.3-2.0~keV) and hard band (2.0-7.0~keV). The limit of
7.0~keV was set in order to minimize contamination by the
particle background, while the boundary of 2.0~keV was chosen in order
to limit the contribution of the diffuse emission (which is mainly
thermal with $\rm{kT=0.6~keV}$) only in the soft band. The source detection was performed 
 in scales of 1, 2, 8 and 16 pixels and a
limiting probability of $\sim1$ chance detection over this
field.  We found a total of 62 sources within the S3 CCD.
Of those, 45 sources are within the D25 radius of NGC~4261,  40 of
which have  significance higher than 3$\sigma$ above the local
background. 
 The absorption corrected luminosities of the latter are above the
Eddington limit for an 1.4\msun neutron star, ranging
 from $2.5\times10^{38}$\ergs~  up to $4.4\times10^{39}$\ergs~
 (with two sources above $10^{39}$\ergs) assuming a 
power-law model  with $\Gamma=1.7$ and  Galactic line-of-sight $\rm{H_{I}}$
column density ($5.8\times10^{20}$~\cmsq; Stark \etal 1992).

\subsection{Spatial distribution of X-ray sources}

Figure 1a shows an adaptively smoothed  image of NGC~4261 in the full
band, together with the D25 ellipse of the galaxy and
the active field of the S3 ACIS-CCD. 
For reference in Figure 1b we present a DSS2 R-band image of the
galaxy in the same scale, with the discrete sources marked by the best fit
$3\sigma$ ellipses to the spatial distribution of their photons\footnote{http://asc.harvard.edu/ciao/download/doc/detect\_html\_manual/}.
From the X-ray image is clear that the spatial distribution of the
discrete sources does not follow the distribution of optical
star-light, but shows instead a distinct asymmetric pattern. 
In the following discussion we focus on the sources within
the D25 ellipse. 

 In order to assess the probability that the non-uniform distribution
 may be observed by chance, we performed a Kolmogorov-Smirnov (KS) test
to compare the distribution of the position angles (PA) of the
sources (measured clockwise from  the  
North-South direction), with (a) a uniform distribution and (b) 
 the azimuthal profile of the optical light of the galaxy. 
For the latter we created a sample of ``optical sources'' based on the star-light distribution in the DSS2 R-band image,   by randomly drawing position angles in 2\deg~ sectors from a flat parent
distribution (equivalent to assuming constant surface brightness within each
2\deg~  sector). The number of draws in each sector was proportional to
the fraction of the total galactic light included in it, after
removing any strong point-like optical sources (most probably
associated with foreground stars or background galaxies/AGNs). 
We found that  both comparison distributions are inconsistent 
with  the observed X-ray source azimuthal distribution  above the
99.9\% confidence level (Fig.~2a). This result holds even for model 
distributions with as few as 10 sources.

 A  search for irregularities in the azimuthal distribution of
the PA of the sources, using the Bayesian blocks method (Scargle 1998;
 Scargle \etal 2003, in prep) also shows that there is an excess of
sources between PA=140\deg~ and PA=190\deg~ at a confidence level
greater than 99\% (estimated from the prior of $\gamma=4.5$ used in the
decomposition; however see
Scargle \etal 2003, in prep\footnote{see also http://space.mit.edu/CXC/analysis/SITAR/functions.html}, for a caveat on this interpretation of the 
 value of the prior).
Therefore, we conclude that the spatial
distribution of the X-ray sources is not the same as that of
the optical light tracing the stellar population.

\section{Other examples: NGC~4697} 

 Indication for a different  distribution  of X-ray sources and
optical star-light has been also seen in NGC~720 (Jeltema \etal 2003), 
which has relatively young stellar populations ($\sim5$~Gyrs; 
Trager \etal 2000).
The X-ray sources in NGC~720 appear to trace out
spiral arms, but performing the same analysis as above we confirm that 
this result is not statistically significant (Jeltema \etal 2003).
Based  on the KS test we find that the X-ray source distribution is different
from the optical distribution or a constant at the 50\% and 70\%
confidence levels 
respectively. Similarly, the Bayesian blocks method showed that any
local enhancement of sources is significant only at the 63\%
confidence level. 

   We searched the
\chandra archive for deep observations of 
nearby early-type galaxies, which also did not show any evidence for
fine structure (Schweizer \& Seitzer 1992) or recent merging activity, in order to construct a
comparison sample. We obtained data for  NGC3379, NGC4636 and
NGC4697, the general properties of  which are presented in Table~1.  
 We analyzed the data for these three galaxies in the same way as for
NGC~4261 (\S2.1). 
A KS  test between the X-ray source distribution and the comparison
distributions (calculated as described in \S2.2) showed that only in
the case of NGC~4697 we can rule out at the 98\%  confidence level (Fig.~2b)
that they belong to the same population ($\sim95\%$ probability that
there is a local enhancement of sources based on Bayesian blocks). 
 In the other two cases  (NGC~3379 and NGC~4636) these
confidence levels are $\sim30$\% and $\sim25$\% respectively, not
allowing us to draw any definitive conclusions.

\section{Discussion and Conclusions} 

Our results show evidence for a non-uniform  spatial distribution of the
X-ray source population in two nearby apparently normal elliptical
galaxies.  These galaxies have very low fine-structure parameters
($\Sigma=1$  for NGC~4261 and $\Sigma=0$ NGC4697) indicating 
 that if they are merger products the merger event took place at least a
few Gyrs ago. This is because  merger simulations  indicate that
 ripples, shells or strong tidal tails are not easily
observable
for much longer than $\sim10^{9}$~yr after the  relaxation of the
merged system (e.g. Quinn 1984; Hernquist \& Quinn 1988, 1989).
Moreover,  faint traces of tidal tails 
surviving for much longer might not be easily detectable in optical 
observations because of their very low surface brightness (Mihos
1995).
In fact, optical observations indicate stellar populations as old as 9 
and 15~Gyrs in the nuclear regions of NGC~4697 and NGC~4261
respectively (Trager \etal 2000).

 We propose that one or more localized star-formation events which
occurred  at most a few hundred Myr ago, are
responsible for the discrepancy between the  optical and X-ray
morphology of these two galaxies. According to X-ray binary
formation  models, LMXBs are more susceptible to effects
such as supernova kicks and common envelope phases (Kalogera \&
Webbink, 1998) than HMXBs, leading to HMXB formation efficiencies
 10 to 100 times higher than for LMXBs (Portegies-Zwart \& Verbunt
1996).   
In fig.~3  we plot the  $\rm{L_{X}/L_{B}}$ ratios of
star-forming galaxies from the \einstein  survey of Shapley, Fabbiano
\& Eskridge (2002) together with the   mean $\rm{L_{X}/L_{B}}$ ratio
for the discrete X-ray sources in elliptical galaxies
($\rm{log(L_{X}/L_{B})=-3.05\pm0.4}$\footnote{where $\rm{L_{X}}$ is in 
\ergs, in 
the 0.35-10.0~keV band, $\rm{L_{B}}$ is the total B band luminosity in
erg/s}; Athey 2003). From this plot is clear that even after
accounting for a  50\% contribution from the diffuse emission in
star-forming galaxies (e.g. Zezas \etal 2001), the latter have
systematically higher $\rm{L_{X}/L_{B}}$ ratios than the X-ray binary component
of early type galaxies. This together with the fact that individual HMXBs have significantly lower
X-ray to B-band flux ratios than LMXBs (van Paradijs \& McClintock
1995) supports the picture that HMXBs are forming much more efficiently 
than LMXBs.

 Recently Barnes (2003) suggested that shock-induced star-formation
can explain the star-forming activity observed along the tidal tails
of ``The Mice''(Stockton 1974; de Grijs et al. 2003).  According to this model shock waves  developing
along the tidal tails can compress the neutral gas and trigger
star-formation. This picture is consistent with the tail-like
distribution of the observed X-ray sources in NGC~4261, if
the tidal tails are projected against the body of the merger remnant.
If the age of this young stellar population is less than a few hundred Myr,  
it forms X-ray binaries much more efficiently than the populations in the
relaxed merger remnant. On the other hand its optical emission is  diluted by
the optical light of the much stronger old population making
its detection in the optical band very difficult.  Depending on the strength
of the star-formation, 
this may result in an overall projected spatial  distribution of X-ray sources
 which is inconsistent with that of the star-light. This
star-formation event can be well-approximated by an instantaneous
burst, in which case the young X-ray source populations are expected to decay 
in a few hundred Myr after the passage of the shock.  Therefore,
we estimate that the shock should
propagate with a speed of $\sim100$~\kms in order to cover the
 length of the region we observe X-ray sources (20.5~kpc), within the
timescale of HMXB formation ($\sim200$~Myr). This speed is realistic
for shock waves developing in the interstellar medium.

Alternatively, N-body simulations show that
structures resembling dwarf galaxies can form in the tidal tails as
a merger completes (Barnes \& Hernquist
1992, 1996).  These structures
remain bound to the body of the remnant, orbiting it.
Because the objects formed in this manner have a range of
binding energies in the potential well of the remnant, the
characteristic time for them to fall back onto the remnant can be
much longer than the time for the body of the remnant
to relax (Hernquist \& Spergel 1992; Mihos \& Hernquist 1996).  Thus, depending on the
orbital distribution, the remnant will
eventually lose evidence for a merger, while tidal dwarfs
continue to orbit at large radii.
The interaction between the body of the merger and the condensations
in the tidal tails can trigger small scale star-formation events and
locally enrich the galaxy with a young stellar population. 
 
As was mentioned earlier, a young stellar 
population can form  X-ray binaries very efficiently, while it is very
difficult to detect it against the much brighter stellar populations
of the merger remnant.  
Therefore, depending on the strength
of the star-formation,  this may result in an overall spatial  distribution of X-ray sources
 which is inconsistent with that of the star-light.
 Since fall-back is estimated to
occur for several Gyrs after relaxation,  we expect that even
elliptical galaxies with stellar populations of $\sim10$~Gyrs  may
exhibit non-uniform X-ray source distribution, if they are the
end-points of galaxy mergers.

 Although the spatial distribution of the sources in NGC~4261
indicates that they are associated with the tidal tails, probably 
 both mechanisms can produce populations of numerous, young
X-ray binaries in merger remnants. 
 Given the timescales of the orbits of dwarf galaxies (up to ~ 1 Gyr) 
or the velocities of the shocks along the tidal tails,
 and the lifetime of HMXBs (up to $\sim100$~Myr), we would expect long
periods during which these galaxies have 
uniform 
X-ray source distributions.
 These periods are expected
to be longer in the more evolved systems since the dwarf companions
with short orbits are accreted first, which is consistent with our
finding that this phenomenon appears in only a few of the most evolved
galaxies examined. Studies of a  more complete sample of elliptical
galaxies and
merging galaxies in their latest stages of merging
  will allow us to further investigate this hypothesis.

\acknowledgments

We thank the referee (J. Rose) for a very helpful report.  We also
thank J. Barnes for providing results prior to publication, and
E. O'Sullivan and A. Athey for useful discussions on the X-ray
emission of X-ray binaries in elliptical galaxies.  We thank
P. Ratzlaff and J. J. Drake for making available their code for
Bayesian blocks analysis.  This work has been partly supported by NASA
Grants G01-2116X, G01-3150X, and ATP NAG5-12140 and NSF grants ACI
96-19019, AST 98-02568, AST 99-00877, and AST 00-71019.

{}

\eject

\makeatletter
\def\jnl@aj{AJ}
\ifx\revtex@jnl\jnl@aj\let\tablebreak=\nl\fi
\makeatother
\begin{deluxetable}{lcccc}
\tabletypesize{\scriptsize}
\tablecolumns{10}
\tablewidth{0pt}
\tablecaption{Properties of the galaxies}
\tablehead{ 
\colhead{Galaxy} & \colhead{Hubble Type} & \colhead{Distance$^{\dag}$ (Mpc)} &
\colhead{ $\Sigma^{\ddag}$} &  \colhead{Notes}} 
\startdata
NGC 4261 & E2 & 32.5 & 1.0 & dust, quadrupole structure (a) \\
& & & &  age $\sim15$~Gyrs (b) \\ 
NGC 3379 & E1 &  8.1    & 0.0  & \\
NGC 720  & E5 & 28.0 & -   & in group, age 5~Gyrs (b)   \\
NGC 4636 & E0 & 15.0 &  - & complex X-ray morphology (c) \\
NGC 4697 & E6 & 15.9 & 0.0 & age $\sim9$~Gyrs (b) \\
\enddata
\tablenotetext{\dag}{Galaxy distance assuming $\rm{H_{o}=75~km/s/Mpc}$.}
\tablenotetext{\ddag}{Fine structure parameter from Schweizer \& Seitzer, 1992.}
\tablenotetext{~}{References: (a) Colbert, Mulchaey \& Zabludoff,
2001; (b) Trager \etal 2000; (c) Jones \etal 2002.  }
\end{deluxetable}

\newpage

 \begin{figure}
\caption{Left: (a) A full band (0.3-7.0~keV) adaptively smoothed
\chandra image of NGC~4261; the D25 ellipse is shown by the white
ellipse. The white box indicates the active field of the ACIS-S3
chip.  The green ellipses indicate the position of the sources in the
central,  saturated, region of the image. Right: (b)  The R-band DSS2
image of the galaxy with the 
X-ray sources marked by the red ellipses. The D25 ellipse is again
shown by the white ellipse.  }
\end{figure}

\eject

\begin{figure}
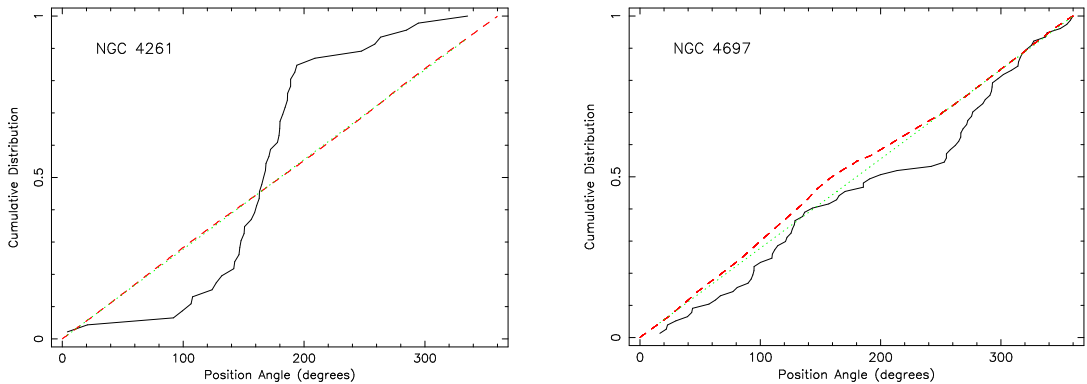

\rotatebox{270}{\includegraphics[width=5.0cm]{f3.ps}}
\hspace{0.75cm}
\rotatebox{270}{\includegraphics[width=5.0cm]{f4.ps}}
\caption{(a) Comparison between the cumulative distribution of the
position angles of the X-ray sources within the D25 ellipse of NGC~4261
 with the azimuthal distribution
of its optical light (dashed red line) and a constant (dotted green line). (b)
The same plot but for NGC~4697.}
\end{figure}

\eject

\begin{figure}
\rotatebox{270}{\includegraphics[width=5.0cm]{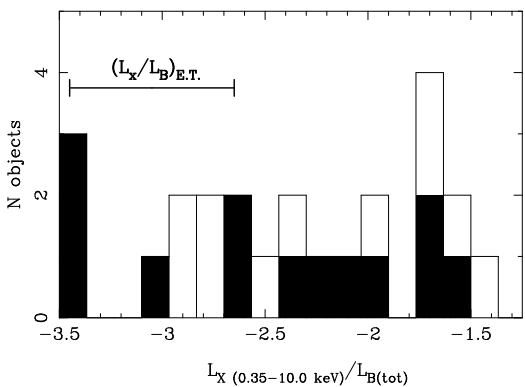}}
\caption{Histogram of the X-ray (0.35-10.- keV) to total B-band
luminosity ratio of spiral galaxies later than Sc type (T$\geq6$) from the \einstein sample
(Shapley \etal 2001). The shaded regions show only the X-ray detections
whereas the unshaded regions show both the detections and non-detections
(upper limits). The horizontal line gives the range of 
$\rm{L_{X}/L_{B}}$ for the discrete source populations of
elliptical galaxies (Athey 2003).  }
\end{figure}

\end{document}